\documentstyle[aps,preprint]{revtex} 

\begin{document}
\begin{center}
{\Large{\bf Flow effects on multifragmentation in the canonical model}}\\

\vskip 1.0cm
S. K. Samaddar$^1$, J. N. De$^{1,2,3}$ and S. Shlomo$^{3}$\\ 
$^1$Saha Institute of Nuclear Physics, 1/AF Bidhannagar, Kolkata 700064, India\\
$^2$Variable Energy Cyclotron Center, 1/AF Bidhannagar, Kolkata 700064, India\\
$^3$The Cyclotron Institute, Texas A$\&$M University, College Station,\\
Texas 77843, USA
\end{center}

\begin{abstract}
A prescription to incorporate the effects of nuclear flow on the process of
multifragmentation of hot nuclei is proposed in an analytically solvable 
canonical model. Flow is simulated by the action of an effective negative
external pressure. It favors sharpening the signatures of liquid-gas phase
transition in finite nuclei with increased multiplicity and a lowered
phase transition temperature.
\end{abstract}
\vskip 1.0cm
PACS Number(s): 24.10.Pa, 25.70.-z, 25.10.Pa
\newpage

In intermediate energy heavy ion reactions, particularly for the central
and near-central collisions, the colliding nuclei get compressed in
the initial phase with subsequent decompression thereby generating
collective flow energy. At energies around 100 $MeV$ per nucleon or above,
large radial collective flow has been observed in many experiments
\cite{jeo,hsi,kun,poc}. Theoretically it has been surmised that collective
expansion has a strong influence on the  fragment multiplicity. In a
hydrodynamical model with site-bond percolation, it has been shown that
compression is very effective \cite{des} in multifragmentation. Such a
conclusion is further reached in microscopic BUU-type formulations \cite{xu}
as well as in a grand canonical thermodynamic calculation \cite{pal1}.
Its crucial importance on the extracted value of the freeze-out density
from yield ratios of fragment isotopes differing by one neutron
\cite{alb,cam} in a statistical fragmentation model was also 
pointed out \cite{shl}. 

Speculations have been made connecting multifragmentation to a liquid-gas
type phase transition in finite nuclear systems (detailed references may be
found in \cite{bon,gro1,bor}). Experimental determination of the caloric
curves in nuclear multifragmentation studies suggest strongly the occurrence
of such a transition. The determination of temperature, however, is still
shrouded in uncertainty and the order of the transition is a
subject of controversy. Theoretical models of different genres have 
been proposed; these include percolation \cite{cam1}, lattice-gas 
\cite{pan,cho}, statistical canonical \cite{bon} and microcanonical models
\cite{gro1} and semi-microscopic models like finite temperature
Thomas-Fermi theory in both nonrelativistic \cite{de} and relativistic
\cite{sil} framework. Many of these models are based on the phase space
considerations though they differ in details. A canonical model based on
this consideration which is analytic in nature has been proposed in
Ref.\cite{cha} and some applications \cite{bha,tsa} of this model have been
made in the context of nuclear mutifragmentation. This model is
comparatively easily tractable, but still powerful enough to reproduce many
of the features of nuclear multifragmentation including liquid-gas phase
transition that can be correlated to 
some of the experimental data. This model,
however, does not include the effects of nuclear flow observed in
intermediate energy heavy ion collisions.
In this communication we incorporate nuclear flow in the
model and study its effect on some inclusive multifragmentation
observables. 
 
        The flow effects are simulated through an
external negative pressure \cite{pal1}. In the stationary freeze-out
volume calculation as no nucleonic matter exists beyond the freeze-out
boundary, the external pressure is assumed to be zero.
A positive uniform external
pressure, {\it i.e.} an inwardly directed pressure, gives rise to
compression of the system. Similarly, a negative external pressure gives
rise to an inflationary scenario (as in the case of early universe
\cite{gut}, for example). The expanding nuclear system can then be
simulated as under the action of an effective negative external pressure.
We define the flow pressure to be equal and opposite to this negative
external pressure. It should be pointed out that the validity of the
model depends on the assumption that the thermodynamic equilibration
time is small compared to the time scale for the expansion of the
system. This is expected to be fulfilled \cite{pal1} when
$v_{flow}/\langle v \rangle $ is much small compared to unity; here
$\langle v \rangle $ is the average nucleonic velocity. This limits
the applicability of the model to flow energy  upto
$\sim$5 $MeV$ per nucleon.

We consider an excited nuclear system at a temperature $T$ and under an
external pressure $P$ (negative in our case, the flow pressure
$P_{fl}=-P$). The system consists
of $N$ neutrons and $Z$ protons, the total number of nucleons
being $A (=N+Z)$. The partition function $Q_{A,Z}$ of the system \cite{ma}
is given by
\begin{eqnarray}
Q_{A,Z}&=&exp\left(-G/T\right) \nonumber\\
&&=\sum_r exp\left[\left(E_r + PV_r\right)/T\right].
\end{eqnarray}
Here $G=E-TS+PV$ is the Gibbs potential, $E_r$ the state dependent 
energy and $V_r$ the state dependent
volume. If $\omega_{ij}$ represents the partition function for the fragment
$(ij)$ consisting of $i$ nucleons and $j$ protons, the 
partition function of the system
$(A,Z)$ fragmenting into all possible configurations $\{{\bf n\}}$, assuming
the fragment pieces are non-interacting, is given by
\begin{eqnarray}
Q_{A,Z}=\sum_{\{{\bf n\}}}\prod_{i=1}^A \prod_{j=0}^Z
\frac{\left(\omega_{ij}\right)^{n_{ij}}}{n_{ij}!}.
\end{eqnarray}
Here $n_{ij}$ is the number of $(ij)$ species present. The sum runs over all
possible configurations conserving nucleon number and charge. The average
multiplicity of $(ij)$ species is
\begin{eqnarray}
\langle n_{ij}\rangle=\frac{\omega_{ij}}{Q_{A,Z}}Q_{A-i,Z-j}.
\end{eqnarray}
The function $Q_{A,Z}$ can be easily calculated using the
recursion relation \cite{cha}
\begin{eqnarray}
Q_{A,Z}=\frac{1}{A}\sum_{i=1}^A\sum_{j=0}^Z i\omega_{ij}Q_{A-i,Z-j}.
\end{eqnarray}
The partition function is built up defining $Q_{00}=1$. The partition
function $\omega_{ij}$ is 
\begin{eqnarray}
\omega_{ij}=\sum_k \int \frac {d^3pd^3r}{h^3} exp\left[-\left(E_{ij}^k +
\frac{P_{ij}V}{n_{ij}}\right)/T\right],
\end{eqnarray}
where $P_{ij}$ ($\sum_{ij}P_{ij}=P$) is the partial pressure due to the
$(ij)$ species and
\begin{eqnarray}
E_{ij}^k=\frac{p^2}{2mi}+\epsilon_{ij}^k+V_{ij}^C .
\end{eqnarray}
Here the first term on the right hand side denotes the center of mass
kinetic energy and $\epsilon_{ij}^k$ refers to the energy of the $k$-th
internal state of the fragment; $V_{ij}^C$ is the single-particle Coulomb
energy which we evaluate in the complementary fragment 
approximation \cite{gro2}. Equation (5) reduces to
\begin{eqnarray}
\omega_{ij}=\frac{\left(2\pi mT\right)^{3/2}}{h^3}i^{3/2}q_{ij}\int dV
exp\left(-P_{ij}V/n_{ij}T\right),
\end{eqnarray}
where
\begin{eqnarray}
q_{ij}=\sum_k exp\left[-\left(\epsilon_{ij}^k+V_{ij}^C\right)/T\right].
\end{eqnarray}
We do not have any {\it a priori} notion about the dependence of
$P_{ij}$ and $n_{ij}$ on volume as well as on temperature. We, therefore,
make a simplifying assumption that the dependence of $P_{ij}V/n_{ij} =
P_{ij}/\rho_{ij}$ ($\rho_{ij}$ being the density of the ($ij$) species)
on temperature is linear. It will be seen later that this is tantamount
to assuming the
flow energy of a fragment proportional to temperature. Such a
prescription may not be unjustified as both stronger compression
(hence collective flow) and larger temperature of the fragmenting
system result from enhanced bombarding energy.
We then write $P_{ij}/\rho_{ij} = C_{ij}T$, $C_{ij}$ being
a constant for the fragment species. 

For fragment masses upto $i=16$, the input for $\epsilon_{ij}^k$ is taken
from the experimental data; for fragment masses above 16, the liquid-drop
expression
\begin{eqnarray}
q_{ij}=exp\left[\left(W_0i-\sigma (T)i^{2/3}+
a_iT^2-V_{ij}^C\right)/T\right],
\end{eqnarray}
is taken using Fermi-gas approximation. Here the volume energy term
$W_0$=16 $MeV$, the temperature dependent surface tension is $\sigma (T)
=\sigma_0[(T_c^2-T^2)/(T_c^2+T^2)]^{5/4}$ with $\sigma_0=18$ $MeV$ and the
critical temperature $T_c=18$ $MeV$. The level density parameter 
is taken as $a_i=i/16$ $MeV^{-1}$. 

The total energy of the system is evaluated as
\begin{eqnarray}
E&=&\frac{1}{Q_{A,Z}}\sum_r E_r
exp\left[-\left(E_r+PV_r\right)/T\right]\nonumber\\
&&=\sum_{ij}\langle n_{ij}\rangle\left[\frac{3}{2}T+\left\{i\left(-W_0 +
T^2/16\right)+\sigma (T)i^{2/3}-T\frac{d\sigma}{dT}i^{2/3}+V_{ij}^C
\right\}\right]\nonumber\\
&&-P\langle V \rangle.        
\end{eqnarray}
In deriving Eq.(10), use has been made of the same approximation as in
Eq.(7).
The first term in the square bracket is the kinetic energy of the fragments
for the center of mass motion and the term within the curly bracket is the
internal energy of the fragments lifted by the Coulomb energy. The last term
is identified as the flow energy (note here that $P$ is negative).
In absence of a better prescription, we have replaced the average volume
$\langle V \rangle$ by a freeze-out volume $V_f$. It is then seen that
the flow energy $e_{fl}^{ij}$ of a fragment belonging to the $(ij)$
species is $P_{ij}/\rho_{ij}$. We then have $e_{fl}^{ij}=C_{ij}T$. 
We consider the flow to
be radial. As the heavier fragments are formed relatively closer to the
center, the flow energy per particle decreases with the mass number of the
fragment. So we parametrize $e_{fl}^{ij}$ as $\delta i^{\alpha}T$
with $\alpha<1$. The parameter $\delta$ determines the flow energy 
of a nucleon at a temperature $T$. The total flow energy is 
\begin{eqnarray}
E_{fl}^{tot}=\delta T \sum_{ij}\langle n_{ij} \rangle i^{\alpha}.
\end{eqnarray}
The decrease of flow energy per particle 
with increasing mass of the fragment
is taken care of through the parameter $\alpha$. 
It can be checked that for $\alpha=1$, the fragmentation
pattern remains unaltered. With these prescriptions, the integral 
pertaining to Eq.(7) is 
$V_f exp(\delta i^{\alpha})$. The effect of flow is thus tantamount to
an increase in the effective freeze-out volume which is dependent on the
fragment species. The larger the species, the larger the effective freeze-out
volume. Such an effect was already observed in a previous analysis of 
experimental data with radial flow \cite{shl}.

In order to study the flow effects on nuclear multifragmentation,
results are shown for $^{197}Au$ taken as a representative system
alongwith those for $^{109}Ag$ to explore the mass dependence of
the observables calculated. An {\it ab initio} determination of the
parameters $\alpha $ and $\delta$ is beyond the scope of a statistical
model.  We vary the parameters $\alpha$ and $\delta$ to study their
sensitivity on the observables. In Fig.1, the average per nucleon 
multiplicity $\langle M \rangle /A$ (top panels) and the average 
number of intermediate mass fragments 
per nucleon $\langle N_{IMF} \rangle /A$
(bottom panels) are displayed as a function of temperature. The
$IMF$'s are defined as fragments with $3 \le Z\le 20$. In panel $(a)$
the fragment multiplicities that are displayed are calculated at a
constant freeze-out volume $V_f=6V_0$ where $V_0$ is the normal volume of
the fragmenting system. All the subsequent calculations at constant 
volume are done at the aforementioned $V_f$. The meanings of the different
lines corresponding to variation of $\alpha$ and $\delta$ are displayed
in the legend. Unless specifically mentioned in the legend, the lines
correspond to $^{197}Au$ as the fragmenting system. The comparison of
the dotted line with the full line shows the influence of flow on
the fragment multiplicity. It is evident that flow enhances 
the multiplicity. We note that the multiplicity $\langle M \rangle /A$
has a sudden enhancement at a particular temperature. It will be seen later
that such enhancement also occurs in the heat capacity and entropy at around
this temperature which we identify as a liquid-gas type phase transition
in a finite nucleus. This transition temperature decreases with
increasing flow. At a constant volume, we note that generally multiplicity 
increases with decreasing $\alpha$. The multiplicity and the transition
temperature are weakly dependent on the parameter $\alpha$. Their
dependence on the mass of the fragmenting system is also not very
significant as is evident from the results displayed for $^{109}Ag$
(dashed line) in the figure. The fragment multiplicity at constant
flow pressure $P_{fl}=0.025 MeV fm^{-3}$ is displayed in the panel
$(b)$. The values of the parameter sets corresponding to different lines
are given in the legend. For all the results presented in Figs.1-3,
the legends of panels $(a)$ and $(b)$ apply for calculations performed
at constant volume and at constant flow pressure, respectively. From the
comparison of the solid line and the dotted line it is 
found that the multiplicity
increases significantly with the increase in the flow energy. As in
the case of constant volume, the multiplicity is seen to be not
sensitive to the parameter $\alpha$ and the mass of the fragmenting
system. The jump at the transition temperature is somewhat more 
marked here as compared to that for constant volume calculations. The
total flow energy is quite insensitive to the parameter $\alpha$ and
is mostly governed by the parameter $\delta$. For $\delta=0.5$,
at the transition temperature the flow energy is $\sim 1.6$ $MeV$
per nucleon which increases to $\sim 2.3$  $MeV$ per nucleon for
$\delta$=0.8. In the lower panels of the figure, the average number
of intermediate mass fragments per nucleon 
$\langle N_{IMF} \rangle /A$ are displayed
as a function of temperature both at constant volume and at constant
flow pressure as indicated. Below the transition temperature the
number of $N_{IMF}$'s are very small; at the transition temperature
there is a sudden enhancement in the $IMF$ multiplicity. The
dependence of $\langle N_{IMF} \rangle /A$ on the 
parameters $\alpha$ and $\delta$ as
well as on the mass of the fragmenting system are similar as found
for the fragment multiplicity $\langle M \rangle /A$. Experimentally
the multiplicities are measured as a function of excitation energy.
The calculated results alongwith the measured 
$\langle N_{IMF} \rangle /A$ as a
function of $E^*/A$ both at constant volume and at constant pressure 
are displayed in Fig.2. The average multiplicity per nucleon
$\langle M \rangle /A$ is seen to increase smoothly with $E^*/A$;
the $\langle N_{IMF} \rangle /A$ is found to rise and fall smoothly
as a function of excitation energy. It is found that the calculated
results at constant pressure conforms better with the experimental data.
In the experimental situation, the mass of the fragmenting system
decreases appreciably with the excitation energy. However, from the
calculated results for $Ag$ and $Au$, we find that the $IMF$ 
multiplicities nicely scale with the mass of the fragmenting system.
This justifies the comparison of $\langle N_{IMF} \rangle /A$ calculated
for a single system for all the excitation energies with the 
experimental data.

The caloric curves, {\it i.e.} the dependence of the excitation energy
on temperature both at constant volume (top panel) and at constant 
pressure (bottom panel) are presented in Fig.3. 
The dashed line corresponds to $^{109}Ag$, the other ones refer to
$^{197}Au$ with different choices of parameters as explained in connection
with Fig.1. 
 The caloric curve at constant volume shows a monotonic
increase of temperature with excitation energy; however, a clear plateau is
observed at around $T$=6.7 $MeV$ for calculation without flow and at $\sim$ 5.8
$MeV$ for all values of $\alpha$ chosen with 
$\delta$= 0.5. A few representative experimental data
(given by filled circles \cite{poc} and open triangles \cite{hau})
 are shown in the figure. There is a wide variation in mass of the excited 
fragmenting system in these data. Mass variation is an important factor
that has been often emphasized \cite{nat1} in any interpretation 
of the caloric curve; however, in the mass range 100-200, there is 
not much quantitative change in the experimental data \cite{nat2}. This is 
also reflected in our calculations.  
  It is seen that with a
modest flow energy of $\sim 2$ $MeV$ per nucleon around the transition
temperature, the qualitative features of the data can be 
fairly reproduced.  The
caloric curve at constant flow pressure, on the other hand, exhibits
instead of a plateau a mild undulation in a very narrow region of 
temperature near the phase transition. The excitation energy is triple
valued at a fixed temperature in this region. This corresponds to three
different freeze-out volumes. (For figures 1 and 5, the relevant quantities
are taken at the highest volume where $G$ is found to be the minimum.)
In a canonical model without flow, such a behavior has also been observed
at constant thermal pressure by Das {\it et al} \cite{das}. Inspection 
of the caloric curves both at constant volume and at constant pressure
shows that they are nearly insensitive to the values of $\alpha$ and 
the mass of the fragmenting systems chosen. However, increase in flow
energy (increase in $\delta$) reduces the transition temperature.

The heat capacity at constant volume $C_v$ 
as a function of excitation energy is shown in the top 
panel of Fig.4 for the system $^{197}Au$ with $\alpha$ =0.8
and values of $\delta$ as marked in the figure. The peaked structure in $C_v$ 
signals a liquid-gas phase transition, the peak becoming stronger with
increasing flow. Results corresponding to the choice of other parameters
are not shown as they yield very similar results.  
The heat capacity at constant flow pressure (bottom panel) with $\delta$
=0.5 and $\alpha$ =0.8 shows a negative branch in the excitation energy zone 
corresponding to the narrow temperature range where the caloric curve
displays a negative slope in the undulating region. The dashed vertical
lines correspond to the maximum and minimum in the caloric curve where
$C_p$ is discontinuous. Similar behavior has also been observed in
the lattice-gas model by Chomaz {\it et al} \cite{cho}.
The qualitative nature of $C_p$ with choice of other flow parameters
remains unchanged and are not shown.

The entropy per particle $S/A$ as a function of temperature at constant
volume and at constant flow pressure $P_{fl}$ are dispalyed in the top and
in the bottom panel of Fig.5, respectively for the values of the flow parameters
as given in the figure.
At the transition temperature,
there is a jump in the entropy which becomes more pronounced for
calculations at constant $P_{fl}$. The larger entropy at any particular
temperature with flow can be understood either from the enhanced
fragment multiplicity with flow or from the increased effective
freeze-out volume.

In summary, we have performed calculations for multifragmentation of a
heated nucleus in a canonical model with incorporation of flow both
at constant volume as well as at constant flow pressure. It may be
pointed out that under the experimental conditions none of these
constraints may exist. In the absence of any definite knowledge
of the actual scenario, the calculations are done with these
constraints imposed. It is found that the average multiplicity increases with 
flow; the average IMF multiplicity shows a rise and fall with excitations
commensurate with the experimental data. The calculated caloric curves also
follow the experimental trend very closely. The plateau in the caloric curve
and the peaked structure of the  corresponding heat capacity at around
5-6 $MeV$ signal a liquid-gas phase transition in the finite nuclear
systems. At constant flow pressure, the caloric curve shows a negative
slope in a small domain of temperature and gives rise to negative heat
capacity. Negative heat capacity at constant thermal pressure has been
observed in the same model without flow \cite{das}; it is interpreted
as arising in regions of mechanical instability where the isobaric
volume expansion coefficient is negative. The same effect is seen to
persist with incorporation of flow. A sudden jump in entropy is also
seen, both at constant volume and at constant pressure. It is interesting
to note that the maximum in the $\langle N_{IMF} \rangle$, the peak
in $C_v$, the discontinuity in $C_p$ and the sudden jump in entropy
are all around the same temperature signalling a liquid-gas phase
transition.

This work was supported in part by the U.S. Department of Energy under
 Grant No.  FG03-93ER40773. J.N.D. gratefully
 acknowledges the warm hospitality at the Cyclotron Institute,
 Texas A $\&$ M University.  S.K.S. is very thankful to the Council
of Scientific and Industrial Research of the Government of India,
for the financial support.

\newpage

\centerline {\bf Figure Captions}

\begin{itemize}

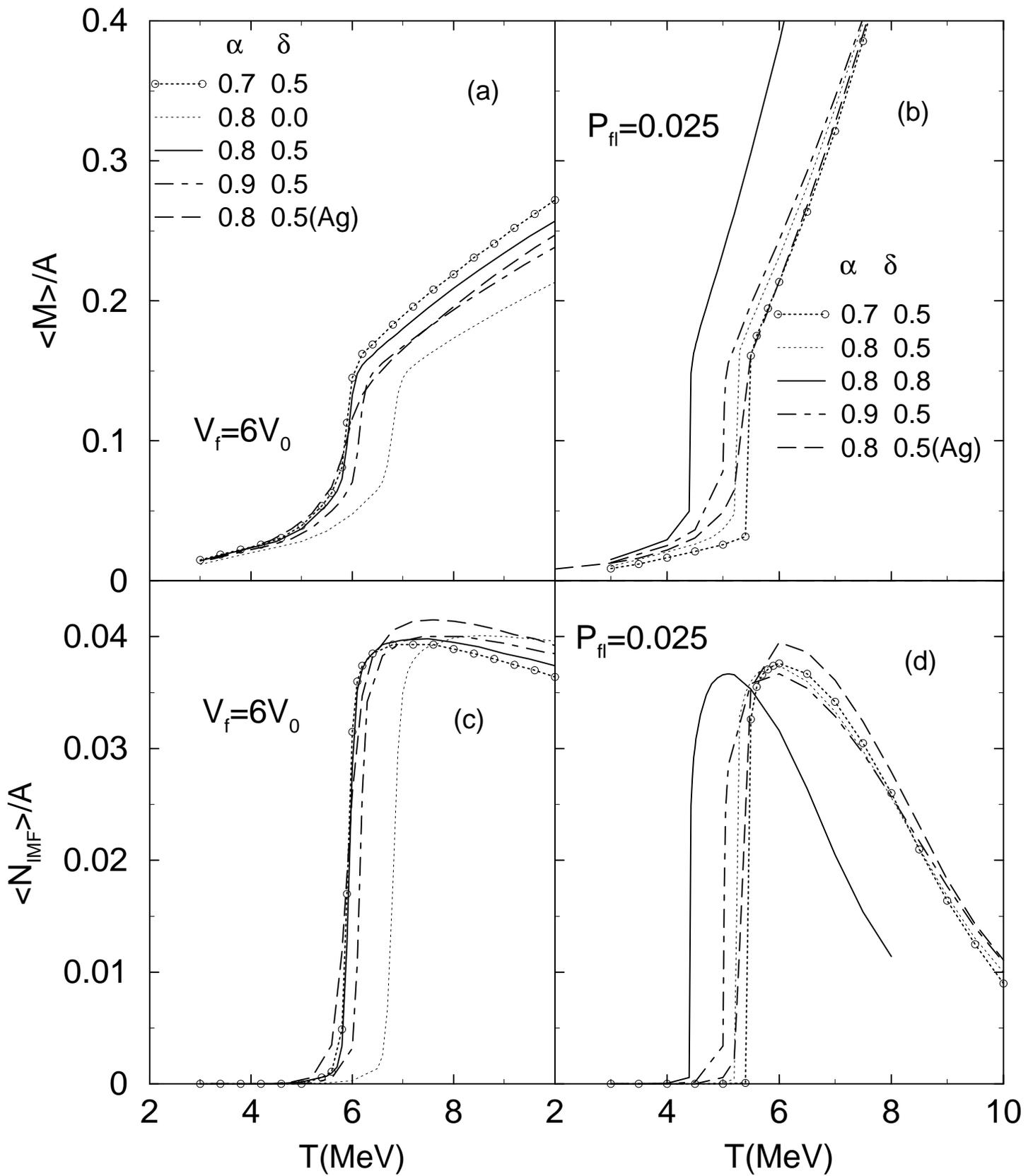
\item[Fig.\ 1] In the top panel the average 
multiplicities per nucleon $\langle M
\rangle /A$ as a function of temperature at constant volume $(a)$
and at constant flow pressure $P_{fl}$=0.025 $MeV fm^{-3}$ $(b)$
are shown. All lines correspond to $^{197}Au$ except the dashed line
that refers to $^{109}Ag$. The different lines refer to different sets of
flow parameters as given in the legend.
In the bottom panel the average $IMF$ multiplicity per nucleon
$\langle N_{IMF} \rangle /A$ is displayed both at constant
volume $(c)$ and at constant flow pressure $(d)$.
The notations for the panels $(c)$ and $(d)$ are the same as those in
the panels $(a)$ and $(b)$, respectively.

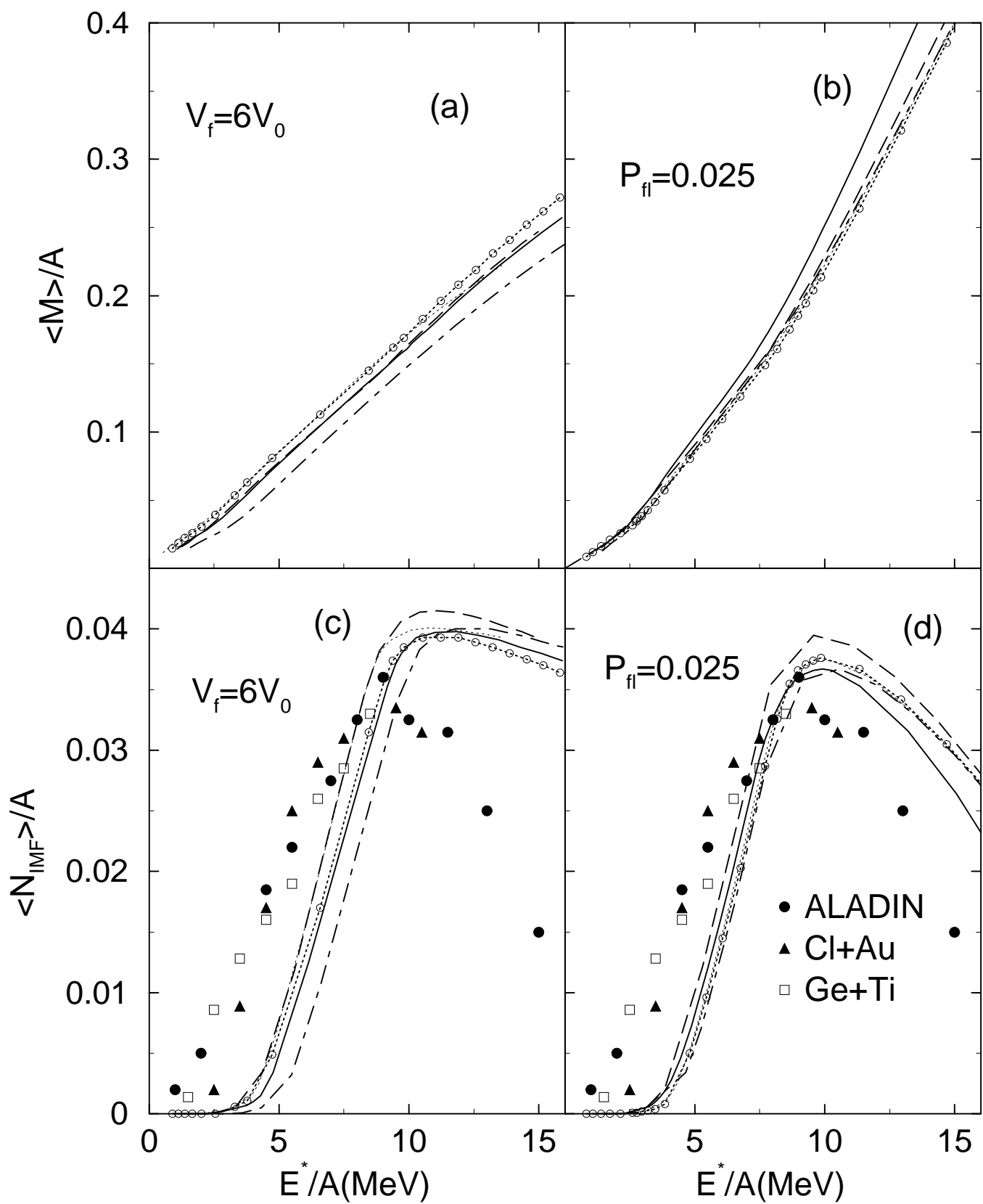
\item[Fig.\ 2] The average multiplicity per nucleon 
$\langle M \rangle /A$ (top panel) and the average
IMF multiplicity per nucleon $\langle N_{IMF} \rangle /A$ (bottom panel) 
are shown as a function of excitation energy. The notations are the same
as in Fig. 1.  Some representative
experimental data for IMF multiplicity are also displayed.

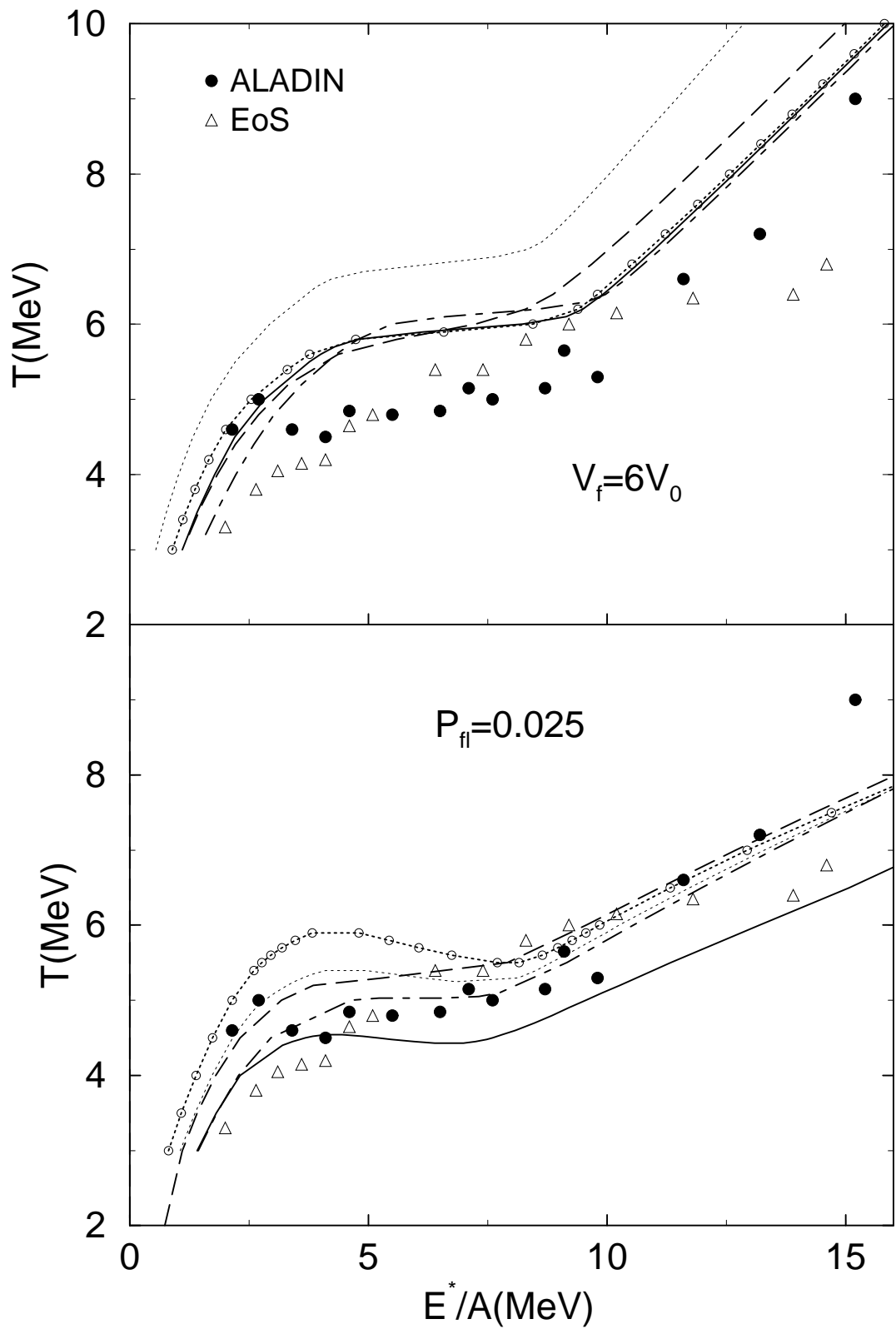
\item[Fig.\ 3] The caloric curves at constant freeze-out volume $V_f=6V_0$
(top panel) and at
constant flow pressure $P_{fl}=0.025$ $MeV$ $fm^{-3}$ (bottom panel). 
The notations are the same as in Fig.1.
The experimental data refer to \cite{poc} (filled circles) and \cite{hau}
(open triangles).

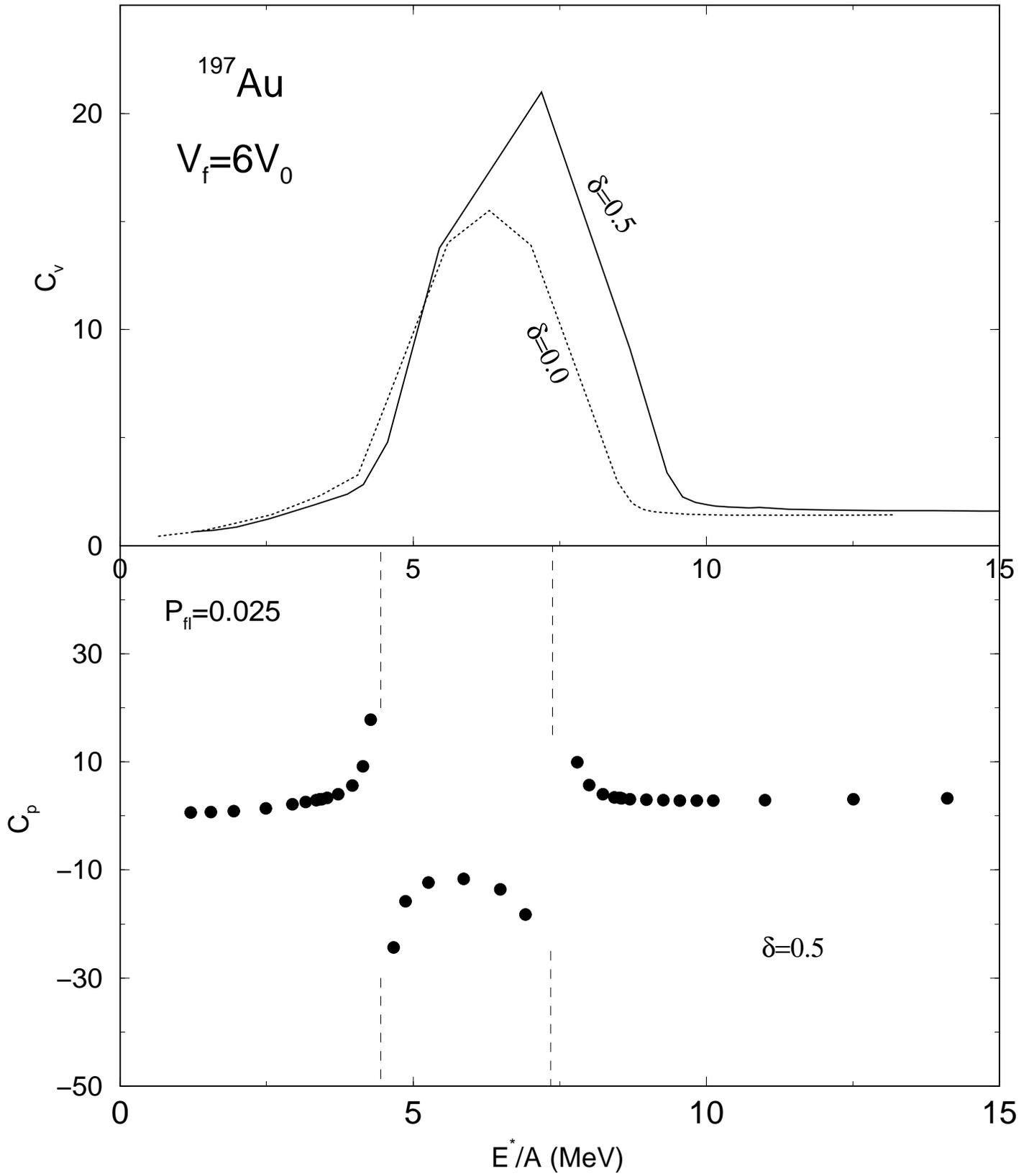
\item[Fig.\ 4] The heat capacity per nucleon at constant volume $6V_0$
(top panel) and at constant flow pressure 0.025 $MeV$ $fm^{-3}$
(bottom panel) are displayed for $^{197}Au$ with $\alpha$=0.8 and
$\delta$ as indicated.
The meaning of the vertical
dashed lines is explained in the text.

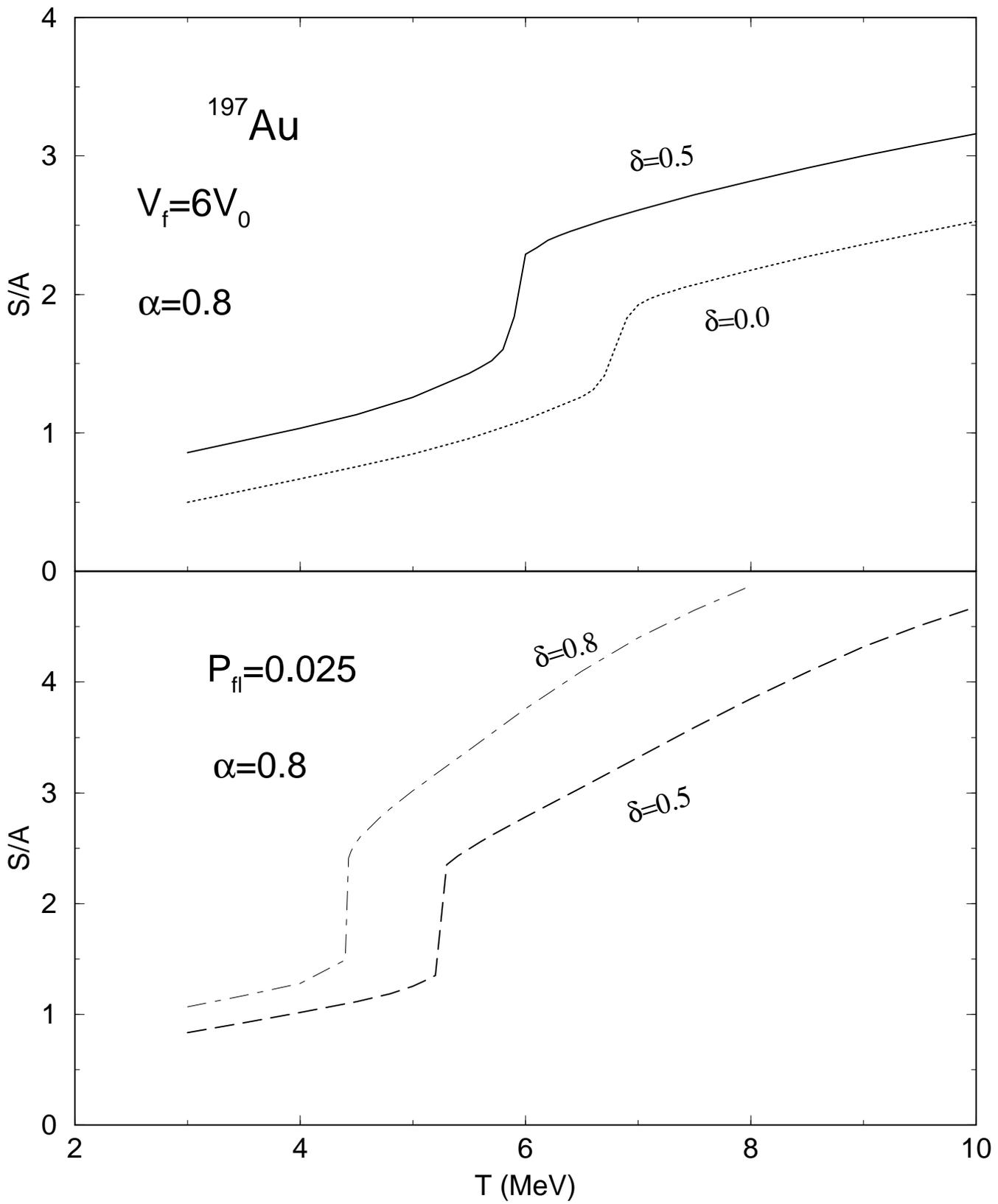
\item[Fig.\ 5] The entropy per nucleon $S/A$ at constant volume $6V_0$
(top panel) and at constant flow pressure 0.025 $MeV$ $fm^{-3}$
(bottom panel) with $\alpha$ and $\delta$ as indicated 
for the system $^{197}Au$.

\end{itemize}
\end{document}